\documentclass[
    ,final            
  ]
  {aipproc}
\layoutstyle{6x9}
\usepackage{graphicx,amsmath,wrapfig}

\begin{document}
\title{A possible nuclear effect \\ on the NuTeV $\sin^2 \theta_W$ anomaly}


\author{M. Hirai}{
  address={Institute of Particle and Nuclear Studies, 
          High Energy Accelerator Research Organization (KEK)},
   email={mhirai@post.kek.jp}}
   
\author{S. Kumano}{
  address={Institute of Particle and Nuclear Studies, 
          High Energy Accelerator Research Organization (KEK)},
  altaddress={Department of Particle and Nuclear Studies,
           The Graduate University for Advanced Studies \\
           1-1, Ooho, Tsukuba, Ibaraki, 305-0801, Japan},
   email={shunzo.kumano@kek.jp}}

\author{T.-H. Nagai}{
  address={Department of Particle and Nuclear Studies,
           The Graduate University for Advanced Studies \\
           1-1, Ooho, Tsukuba, Ibaraki, 305-0801, Japan},
   email={tnagai@post.kek.jp}}

\begin{abstract}
We investigate a possible explanation for the NuTeV anomaly in terms of
a nuclear correction difference between $u_v$ and $d_v$ distributions.
Analyzing deep elastic scattering and Drell-Yan data for nuclear targets,
we try to determine the correction difference and its effect on the anomaly.
We find that the difference cannot be precisely determined at this stage
due to the lack of data which are sensitive to the difference. Therefore,
it is difficult to draw a solid conclusion about its effect although
the anomaly could be explained, at least partially, by this kind of
nuclear correction.
\end{abstract}
\maketitle

\section{Introduction}

Weak-mixing angle is one of the fundamental constants in the standard model.
Using neutrino and antineutrino deep inelastic scattering data, the NuTeV
collaboration found that their angle 
$\sin^2 \theta_W = 0.2277 \pm 0.0013 \, \text{(stat)} 
                         \pm 0.0009 \, \text{(syst)}$  \cite{nutev02}
in the on-shell scheme is significantly larger than the average of
other measurements, $\sin^2 \theta_W = 0.2227 \pm 0.0004$.
This difference is called ``NuTeV anomaly''. Before discussing any exotic
explanations, we need to investigate possible sources from nucleonic and
nuclear structure. For example, the strange asymmetry $s-\bar s$
and QED effect contribute to the difference. In addition, appropriate
nuclear corrections should be taken into account because the NuTeV 
target is the iron. Among the nuclear corrections, it was pointed out
that the nuclear modification difference between $u_v$ and $d_v$ distributions
could affect the $\sin^2 \theta_W$ determination \cite{sk02}. 
Then, this nuclear modification is determined by using world nuclear-correction
data on the structure function $F_2$ and the Drell-Yan cross section in order
to find the possible origin of the NuTeV anomaly \cite{hkn05}.
This analysis was done by using a $\chi^2$ analysis technique developed
for determining nuclear parton distribution functions \cite{npdf04}.
The following discussions are based mainly on the works in Ref. \cite {hkn05}.

\section{Global analysis for determining nuclear correction difference 
         between $u_v$ and $d_v$}

From the neutrino and antineutrino deep inelastic scattering data, it is
possible to obtain the weak mixing angle by the Paschos-Wolfenstein relation,
$R^-  = ( \sigma_{NC}^{\nu N}  - \sigma_{NC}^{\bar\nu N} ) /
        ( \sigma_{CC}^{\nu N}  - \sigma_{CC}^{\bar\nu N} )
        =  1/2 - \sin^2 \theta_W $,
where $\sigma_{CC}^{\nu N}$ and $\sigma_{NC}^{\nu N}$ are
charged-current (CC) and neutral-current (NC) cross sections, respectively.
Various correction factors to this relation are discussed in Ref. \cite{sk02}.
Among them, we investigate a possible nuclear correction difference 
between $u_v$ and $d_v$. These valence-quark distributions should satisfy
the baryon-number and charge conservations so that they have certain
restrictions \cite{sk02}. In principle, they could be determined by
a global analysis of world data on nuclear structure functions.

Valence-quark distributions are defined in nuclei as \cite{npdf04}
\begin{align}
u_v^A (x,Q^2) & = w_{u_v} (x,Q^2,A,Z) \, \frac{Z u_v (x,Q^2) 
                                         + N d_v (x,Q^2)}{A},
\nonumber \\
d_v^A (x,Q^2) & = w_{d_v} (x,Q^2,A,Z) \, \frac{Z d_v (x,Q^2) 
                                         + N u_v (x,Q^2)}{A},
\label{eqn:wval}
\end{align}
where $w_{u_v}$ and $w_{d_v}$ indicate nuclear modification factors.
It should be noted that the modification factors $w_i$ are
used at any $Q^2$ points in this work \cite{sk02,hkn05}
although they are defined only at $Q^2$=1 GeV$^2$ in Ref. \cite{npdf04}.
The difference is defined $\Delta w_v \equiv w_{u_v} - w_{d_v}$,
and it is expressed by four parameters, $a_v'$, $b_v'$, $c_v'$, and $d_v'$
at $Q^2$=1 GeV$^2$ ($\equiv Q_0^2$):
\begin{equation}
\Delta w_v (x,Q_0^2,A,Z) = \left( 1 - \frac{1}{A^{1/3}} \right)
   \frac{a_v' (A,Z) +b_v' x+c_v' x^2 +d_v' x^3}{(1-x)^{\beta_v}},
\end{equation}
where the value in Ref. \cite{npdf04} is used for $\beta_v$.
The parameters are determined by a global analysis by minimizing
the total $\chi^2$:
$\chi^2 = \sum_j (R_{j}^{data}-R_{j}^{theo})^2 / (\sigma_j^{data})^2$,
where $R_j$ indicates the structure-function and Drell-Yan cross-section
ratios, $F_2^A/F_2^{A'}$ and $\sigma_{DY}^{pA}/\sigma_{DY}^{pA'}$.
The experimental error $(\sigma_j^{data})^2$ is given by the quadratic
summation of statistical and systematic errors.
Uncertainty of the obtained $\Delta w_v$ is calculated
by the Hessian method.

\section{Results}

\begin{wrapfigure}{r}{0.44\textwidth}
\begin{center}
\vspace{-0.1cm}
\includegraphics[width=0.40\textwidth]{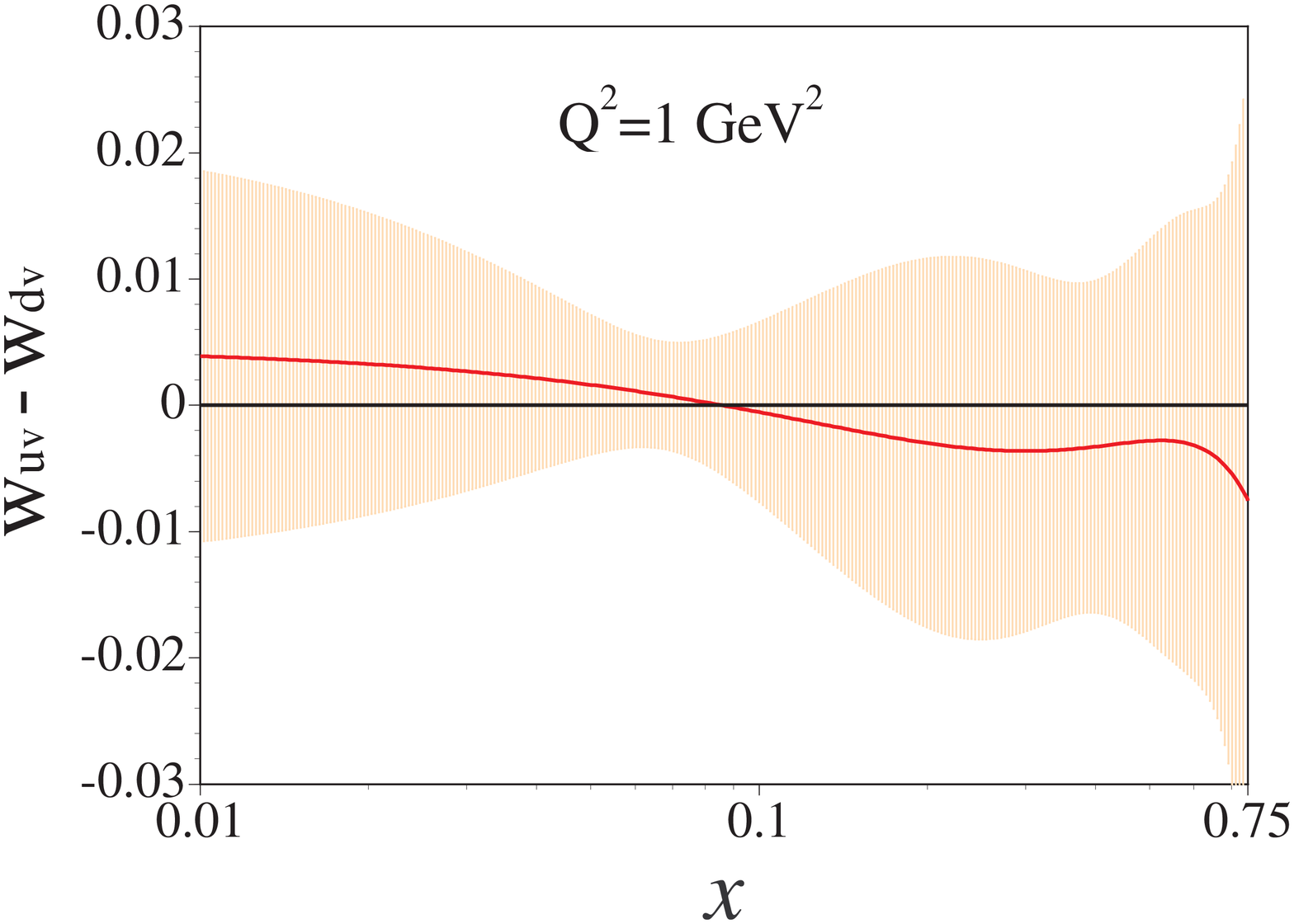}
\vspace{-0.4cm}
  \begin{minipage}{0.37\textwidth}
  \vspace{+0.1cm}
  {\footnotesize {\bf FIGURE 1.}
   Obtained $\Delta w_v$ and its uncertainty at $Q^2$=1 GeV$^2$
   \cite{npdf04}.}
  \end{minipage}
\vspace{+0.2cm}
\end{center}
\end{wrapfigure}
The nuclear valence-quark distributions are defined at $Q^2$=1 GeV$^2$
with the parameters in Eq. (\ref{eqn:wval}). We use the antiquark and
gluon distributions which are determined in \cite{npdf04}. The nucleonic
distributions on the right-hand sides of Eq. (\ref{eqn:wval}) are taken
from the MRST01 distributions.
The distributions are evolved to the experimental points of 
$F_2^A/F_2^{A'}$ and $\sigma_{DY}^{pA}/\sigma_{DY}^{pA'}$ by
the DGLAP evolution equations. Then, the parameters are determined
in comparison with the data so as to minimize the total $\chi^2$.
The obtained optimum distribution at $Q^2$=1 GeV$^2$ is shown by
the solid curve in Fig. 1, and the uncertainty of $\Delta w_v$ is shown
by the band. We find that the difference $\Delta w_v$ is a small quantity.
In fact, the nuclear modifications for $u_v$ and $d_v$ are
assumed to be the same in most theoretical calculations of nuclear
structure functions. The difference could be related to QED effects
on the nuclear parton distributions. However, its precise determination
is not possible from experimental data because the uncertainty
is an order of magnitude larger than the obtained distribution
in Fig. 1.

Because the Paschos-Wolfenstein relation was not directly used 
in the NuTeV analysis, we need to take a weighted average over
the NuTeV kinematics for discussing the anomaly. For example,
there are few data in the large-$x$ region. Fortunately,
such a calculation method is provided in Ref. \cite{nutev02}.
Because the NuTeV definition of the nuclear parton distributions is
slightly different from ours, their relations should be found.
According to the NuTeV convention, the nuclear distributions are
defined as
$x u_v^A = (Z u_{vp}^* +N u_{vn}^*) / A$ and
$x d_v^A = (Z d_{vp}^* +N d_{vn}^*) / A$,
where the NuTeV distributions are denoted with the asterisk (*).
Comparing these expressions with Eq. (\ref{eqn:wval}), we find that 
the modification difference $\Delta w_v$ corresponds to isospin-violating
distributions of the NuTeV collaboration:
\begin{equation}
\delta u_v^* \equiv u_{vp}^*-d_{vn}^* = + \Delta w_v \, x \, u_v ,
\ \ \ 
\delta d_v^* \equiv d_{vp}^*-u_{vn}^* = - \Delta w_v \, x \, d_v .
\label{eqn:uvdv-iso}
\end{equation}
Information is given in Ref. \cite{nutev02} for calculating
these isospin-violating distributions on the weak-mixing angle:
\begin{equation}
\Delta (\sin^2 \theta_W) = 
- \int dx \, \big\{ \, F[\delta u_v^*,x] \, \delta u_v^* (x)
            + F[\delta d_v^*,x] \, \delta d_v^* (x) \, \big\} ,
\label{eqn:del-sinth}
\end{equation}
by using the provided functionals $F[\delta u_v^*,x]$
and $F[\delta d_v^*,x]$.

Using the distribution $\Delta w_v$ together with
Eqs. (\ref{eqn:uvdv-iso}) and (\ref{eqn:del-sinth}), we obtain
$\Delta (\sin^2 \theta_W) = 0.0004 \pm 0.0015$. It should be noted that
the distribution  $\Delta w_v$ is calculated at $Q^2$=20 GeV$^2$,
which is approximately the average $Q^2$ value of the NuTeV experiment.
The magnitude is not large enough to explain the anomaly (0.0050).
However, it does not mean that the considered nuclear modification
mechanism should be ruled out. There is a possibility that the error is
underestimated because the error correlation effects with antiquark
and gluon distributions are not taken into account. In any case,
it is obvious from Fig. 1 that $\Delta w_v$ cannot be determined
at this stage, so that a precise numerical estimation is not possible.
Future efforts are needed for clarifying the nuclear modification
difference $\Delta w_v$ and its effect on the weak-mixing angle
measurement with a nuclear target.

\begin{theacknowledgments}
S.K. was supported by the Grant-in-Aid for Scientific Research from
the Japanese Ministry of Education, Culture, Sports, Science, and Technology. 
\end{theacknowledgments}



\end{document}